\documentclass[letter, 10pt, conference]{ieeeconf}
\IEEEoverridecommandlockouts

\usepackage{cite}

\usepackage{amsmath,amssymb,amsfonts}
\usepackage{algorithmic}
\usepackage{graphicx}
\usepackage{subcaption}
\usepackage{pdfpages}
\usepackage{textcomp}
\usepackage{xcolor}
\usepackage{bm}
\usepackage{booktabs}
\usepackage{svg}
\svgsetup{inkscapelatex=false}
\usepackage{tikz}
\usepackage{hyperref}
\def\BibTeX{{\rm B\kern-.05em{\sc i\kern-.025em b}\kern-.08em
    T\kern-.1667em\lower.7ex\hbox{E}\kern-.125emX}}
\usetikzlibrary{arrows,decorations.pathmorphing,backgrounds,fit,positioning,calc,shapes}

\usepackage[linesnumbered,ruled,algo2e,resetcount]{algorithm2e}%for the environment
\SetAlFnt{\small}
\SetAlCapFnt{\small}
\SetAlCapNameFnt{\small}

\begin{document}
\bstctlcite{MyBSTcontrol}
% --------------------------------------------------------------------------
%% PAPER METADATA

%% TITLE
\title{\LARGE \bf Soothing Sensations: Enhancing Interactions with a Socially Assistive Robot through Vibrotactile Heartbeats
\thanks{Research supported by the UKRI Centre for Doctoral Training in Socially Intelligent Artificial Agents, Grant Number EP/S02266X/1}
}

%Authors metadata
\author{\authorblockN{
Jacqueline Borgstedt$^{a}$,
Shaun Macdonald$^{a}$,
Karola Marky$^{b}$,
Frank E. Pollick$^{a}$,
Stephen A. Brewster$^{a}$
}
\thanks{$^{a}$University of Glasgow, Scotland}
\thanks{$^{b}$Ruhr-University Bochum, Germany}
}

\maketitle

%% ABSTRACT
\begin{abstract}
Physical interactions with socially assistive robots (SARs) positively affect user wellbeing. However, haptic experiences when touching a SAR are typically limited to perceiving the robot’s movements or shell texture, while other modalities that could enhance the touch experience with the robot, such as vibrotactile stimulation, are under-explored. In this exploratory qualitative study, we investigate the potential of enhancing human interaction with the PARO robot through vibrotactile heartbeats, with the goal to regulate subjective wellbeing during stressful situations. We conducted in-depth one-on-one interviews with 30~participants, who watched three horror movie clips alone, with PARO, and with a PARO that displayed a vibrotactile heartbeat. Our findings show that PARO’s presence and its interactive capabilities can help users regulate emotions through attentional redeployment from a stressor toward the robot. The vibrotactile heartbeat further reinforced PARO’s physical and social presence, enhancing the socio-emotional support provided by the robot and its perceived life-likeness. We discuss the impact of individual differences in user experience and implications for the future design of life-like vibrotactile stimulation for SARs.
\end{abstract}

% --------------------------------------------------------------------------
%% PAPER TEXT
\section{Introduction}

The negative effects of stress on wellbeing are widely recognised, with research suggesting that continual exposure to stressors can lead to the development and exacerbation of various psychological and physiological disorders, including anxiety, depression, elevated blood pressure, and hypertension~\cite{Oconnor2021,Shields2017}. Social touch with friends and family can play a pivotal role in mitigating the adverse outcomes of stress exposure: physical interactions with loved ones can reduce subjective stress levels during anxiety-inducing situations~\cite{Jakubiak2019, Coan2006} and enhance stress resilience~\cite{Dagnino-Subiabre2022}. The benefits of social touch extend to other species, such as canines, with a large body of research demonstrating that physical interaction with animals is associated with reduced heart rate, blood pressure, and cortisol levels during stress-inducing situations~\cite{Allen2003, Nimer2007}. Despite the positive impact of social touch on wellbeing, one-in-ten adults report feeling lonely or socially isolated, thus lacking access to social touch. Furthermore, concerns such as allergies or limited financial resources can reduce accessibility to companion animals as a suitable alternative. 

Interaction with socially assistive robots (SARs) has been proposed as a technological alternative for those that cannot access touch-based interaction with other humans or animals~\cite{Willemse2019}. Empirical studies have demonstrated that animal-like (zoomorphic) SARs, such as PARO\footnote{\url{http://www.parorobots.com/}}~\cite{Geva2020,Wada2005,Aminuddin2017}, or the Huggable\textsuperscript{\texttrademark}\footnote{\url{https://robots.media.mit.edu/portfolio/huggable/}}~\cite{Jeong2015}, can provide socio-emotional support through physical interactions. While research indicates touch-based interactions with SARs can benefit dementia patients or children, their interactive capabilities are often simplistic and limited, necessitating novel interaction techniques to develop fully interactive, therapeutic companion robots, suitable for a broader user base\cite{sefidgar2015design, Katsuno2022}. Drawing inspiration from the established benefits of Animal-Assisted Therapy~\cite{Nimer2007}, Sefidgar et al.~\cite{sefidgar2015design} highlight the potential role of enhancing affective touch capabilities of therapeutic robots to enhance their usability and efficacy. 

A small body of research has begun to explore the use of novel haptic cues to enhance the life-like qualities of robots, diversifying the touch-based output of SARs. For instance, Yoshida et al.~\cite{Yoshida2018} developed the BREAR robot, simulating bio-physiological cues like breathing, heartbeat, and temperature, which enhanced the robot's life-likeness and enabled users to infer the its emotional states. Yohanan et al.'s Haptic Creature~\cite{Yohanan2011role} used vibrotactile purring, movable ears, and inflatable lungs to express life-like emotions. Other studies have explored the use of temperature to enhance the perception and interaction with SARs~\cite{Nie2012,Borgstedt2022,Park2014}. Bradwell et al.~\cite{Bradwell2021} provided preliminary evidence that older adults preferred social robots, which exhibit bio-physiological signals such as purring and breathing. While the studies indicate the potential for broadening the spectrum of human-robot touch interactions to include diverse haptic experiences, the implications of these enhancements on a SAR's capacity to replicate socio-emotional interactions and their impact on user wellbeing and users' emotion regulation capabilities remain uncertain. 

In light of this, we conducted a study to examine the effects of exposure to a vibrotactile heartbeat on user wellbeing in the context of interacting with a social robot during a moderately stressful situation. In this study, we simulated a stressful situation in which participants would benefit from social touch by showing them horror movie clips, akin to Nie et al.\cite{Nie2012}, which they watched either alone, with the PARO robot, or a PARO robot augmented with a vibrotactile heartbeat.We investigated the use of a vibrotactile heartbeat due to promising results in Human-Computer Interaction, demonstrating that presentation of heartbeat-like vibrotactile cues may relax users during stressful situations~\cite{Azevedo2017, shaun2024a} or enhance the benefits of social touch with computer interfaces~\cite{MacDonald2022}. However, HCI studies have only investigated the effect of vibrotactile heartbeats displayed through interfaces like wearables, and it is thus unknown if the positive effect on user wellbeing replicate during interactions with a SAR. In this study, we employed qualitative methodology to understand if and how augmentation of PARO with a vibrotactile heartbeat can benefit user wellbeing during exposure to a stressful situation.

The main contributions of our work are:

\begin{enumerate}
\item Showing that vibrotactile heartbeats can make a SAR more engaging and increase its social presence, enhancing the socio-emotional support provided and making it more suitable to comfort users during stressful situations; 
\item Discussing how the presentation and location of haptic hardware can affect the impact of the vibrotactile stimulation, informing the effective design and implementation of vibrotactile heartbeats;
\item Demonstrating that the PARO robot is suitable to comfort users during a stressful situation through its interactive capabilities and animal-like physical attributes.
\end{enumerate}

\section{Methods} \label{sec:methods}

\subsection{Methodology }
This study featured a~3x1 within-participant design. Each participant watched horror movies without the robot (control condition), with the original PARO robot (PARO condition), and with the robot augmented with a vibrotactile actuator simulating a heartbeat (vibrotactile condition). The order of movie clips and experimental conditions was randomized. 

The primary aim of this study is to delve deeply into users' individual encounters with the PARO robot and vibrotactile heartbeat, and its effect on their subjective wellbeing through qualitative research. We adopt an interpretive and constructive epistemological stance, which holds social reality as subjective and shaped by individual experiences. Due to the subjective nature of perceived wellbeing, we believe that a qualitative approach is best suited to gain an understanding on how interacting with PARO affects the wellbeing in our sample. Using a predefined interview guide\footnote{\url{https://osf.io/cb6gw/?view_only=d0d02e5d1a2c4ff09afce71a1b4d4939/}} we conducted one-on-one, semi-structured interviews with each participant after they completed each condition.

We employed Braun and Clarke's reflexive thematic analysis methodology\cite{Braun2006}, to analyse the interview transcripts. The primary researcher, who conducted the interviews, and a secondary researcher, who had no prior interaction with participants, co-coded the data. Initially, each researcher familiarized themselves with the interview transcripts and then systematically generated initial codes to capture key concepts. After aligning initial coding schemes through discussion, both researchers individually analyzed the codes to group them into main and sub-themes. Through discussion and synthesis, three consistent overarching themes, some with sub-themes, were created based on the data, representing meaningful patterns relevant to the research questions. Descriptive names were assigned to the themes, and a comprehensive summary report was prepared by the primary researcher, as detailed in the results section.

\subsection{Participants}
30 participants (17 women, 8 men, 5 non-disclosed), aged 19 to 40, were recruited from the Glasgow University campus. Exclusions included prior diagnosis of anxiety disorders, PTSD, or phobias, which may increase the likelihood of the participant being adversely affected by the emotion-eliciting stimuli; impairments that would impact the interaction with the robot or videos; individuals who previously interacted with the PARO robot or watched the horror movies displayed in the study in the last 6 months. 

\subsection{Procedure}
After the briefing and signing of consent forms, participants resting heart rate was measured with a polar H10 chest strap for a total of 5 minutes, during which participants sat on a chair and breathed regularly.Following, Azevdeo et al.\cite{Azevedo2017}, the resting heart rate was measured to modulate speed of the vibrotactile heartbeat stimulus, which was presented later in the experiment. Participants were then introduced to the PARO robot and its interaction modalities, followed by 2~minutes of free-form interaction. Participants then completed a total of three experimental blocks. Following, Schäfer et al.~\cite{Schaefer2016}, participants completed a 1~minute breathing exercise prior to exposure to the horror movie clip, to induce feelings of relaxation and calmness. Next, participants watched one of the three movie clips either on their own (control condition), with the PARO robot (PARO condition) or with the augmented PARO displaying vibrotactile heartbeats (vibrotactile condition). In both the PARO and vibrotactile condition, participants were instructed to hold their non-dominant hand on PARO’s back and were allowed to pet or stroke PARO with their dominant hand. In both conditions, PARO was placed on the participants lap, facing the participant. After each experimental block there was a 5~minute break. Once the participants completed all three conditions, the primary researcher conducted a semi-structured one-on-one interview, typically lasting between 10~and 15~minutes. The research procedures were fully approved by the University Ethics Committee.
\begin{figure}[ht]
  \centering
  \includegraphics[width=.9\linewidth]{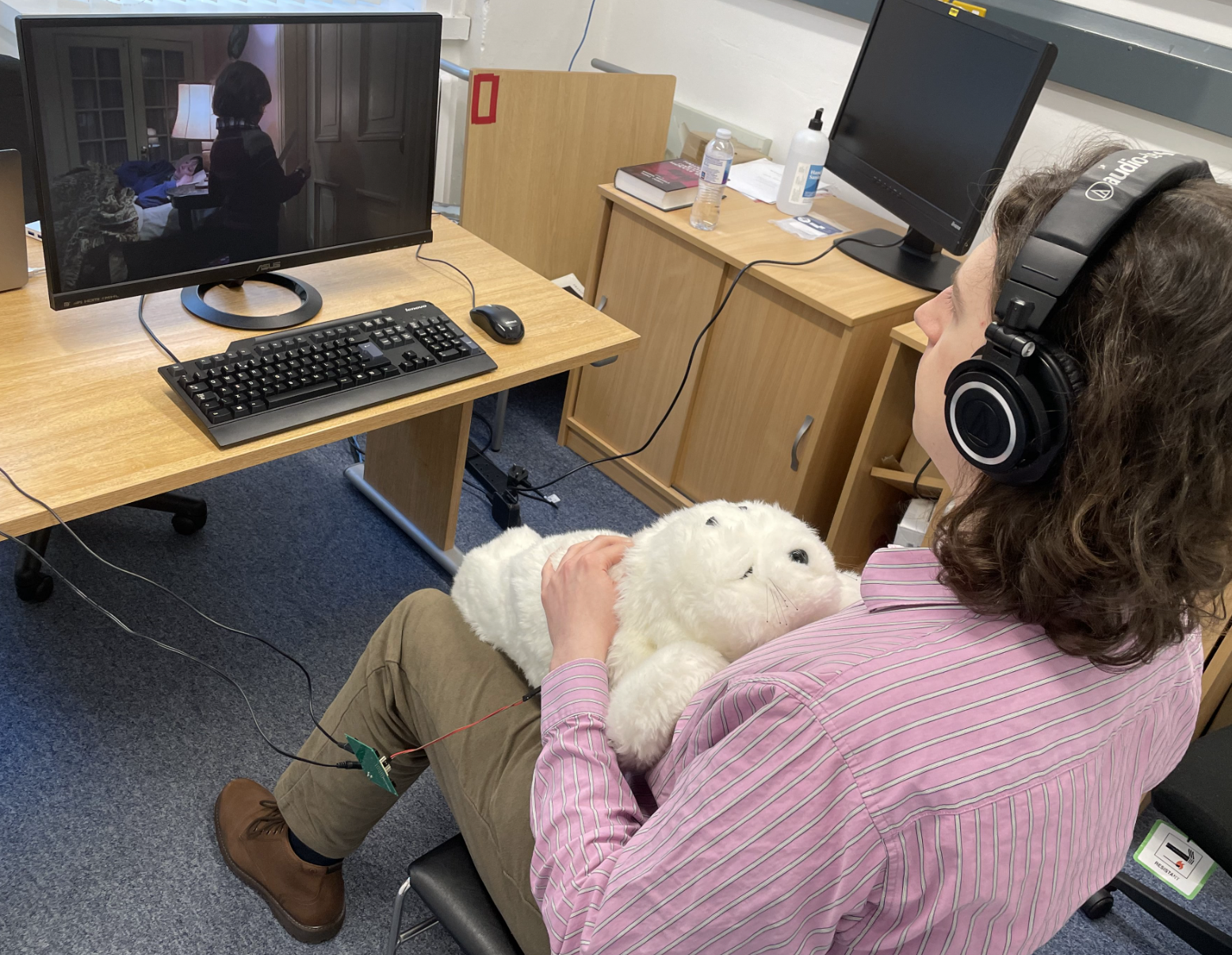}
  \caption{Experimental Setup. Participant watching horror movie with Bluetooth headphones and an 8th generation PARO augmented with a Haptuator Mark II, encapsulated in white fur.}
    \label{fig:hardware}
\end{figure}

\subsection{Robot \& Experimental Set Up }
The robot used in this study was an 8th generation PARO, which resembles a baby harp seal covered in soft white fur (Figure~\ref{fig:hardware}). PARO can sense touch (e.g., petting, hugging etc) and voice inputs (e.g., simple word recognition, voice location) and interacts with the users by making noises and moving its flippers and head. As this study was focused on human-robot touch and investigating the effect of vibrotactile stimulation, PARO’s noise capabilities were disabled. Following~\cite{Yoo2015, MacDonald2022}, the vibrotactile stimuli were delivered using an Haptuator Mark~II\footnote{\url{https://tactilelabs.com/products/hapcoil-one/}} controlled by a Haptu-Amp-Quad board\footnote{\url{https://tactilelabs.com/products/mini-haptuamp/}} connected to an i5 2020 MacBook Pro, with the internal volume set to 50\%. The Haptuator was encapsulated with PlayDough, a soft modelling clay, through which the stimuli could effectively radiate~\cite{MacDonald2022} and covered in white faux fur resembling PARO’s fur. The haptuator was attached to the PARO’s back on the side of the participant’s non-dominant hand via a disguised elastic band (Figure~\ref{fig:hardware}). The movie clips were presented on a 27~inch monitor in an experimental lab, with lights turned off during the movie clips. The lab environment was quiet and over-the-ear Bluetooth headphones were used to deliver the audio from the clips. 

\subsection{Stimuli}
\begin{figure}[ht]
  \centering
  \includegraphics[width=\linewidth, height=1.5cm]{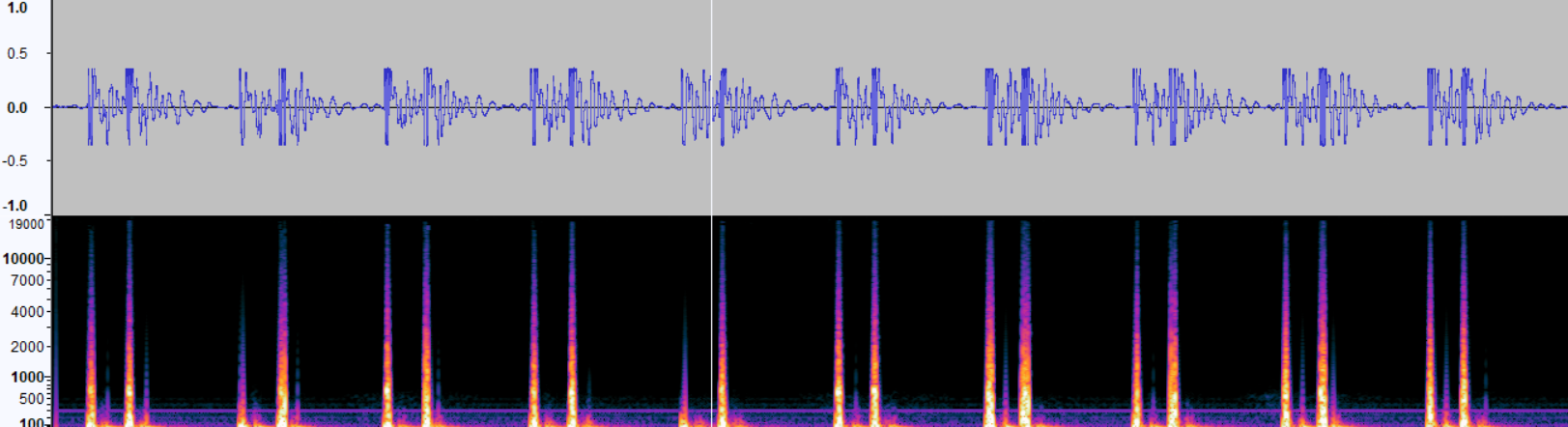}
  \caption{Acoustic wave forms displaying a heartbeat at $\mbox{\textbf{60bpm}}$, which were used to generate the vibrotactile heartbeats.}
    \label{fig:heart}
\end{figure}

We utilised a vibrotactile heartbeat stimulus drawn from MacDonald et al. ~\cite{MacDonald2020}. This was generated using an acoustic recording of a~$60bpm$ heartbeat sourced from an online sound repository \footnote{\url{https://freesound.org/}}, following which volume normalisation of~$89dB$ and a~$300Hz$ low-pass filter were applied (see Figure~\ref{fig:heart} for stimulus waveform and spectrogram). The stimulus had a duration of 10~seconds and was looped. During the study, participants were presented with this heartbeat stimulus. Following Azevdeo et al.~\cite{Azevedo2017} the speed was modulated to~$80\%$ of their own resting heart rate, rounded to the nearest $bpm$. 

The breathing exercise\footnote{\url{https://www.youtube.com/watch?v=uxayUBd6T7M/}} displayed a breathing bubble on screen, which instructed them to inhale, exhale, or hold their breath. The video used was created by Calm\footnote{\url{https://www.calm.com/}}. 

Horror movies have been shown to reliably induce mild to moderate negative affect and stress~\cite{Schaefer2016, Hagenaars2014, mattar2015effect}, and have been used as a means to induce negative affect in empirical studies\cite{Hagenaars2014, mattar2015effect}. Horror movie clips offer an advantage over alternative stress-induction methods, such as the Trier or Mannheim Stress Tasks by allowing repeated exposure, necessary for a within-participant design. We selected clips from three movies \footnote{\url{https://sites.uclouvain.be/ipsp/FilmStim/}}(The Blair Witch Project 1999, Misery 1990, The Shining 1980) from an experimentally validated database for emotion-eliciting films~\cite{Schaefer2016}, which were similar in terms of length (each aproximately 4 mins) and had been demonstrated to induce similiar mean negative affect and anxiety scores. For more information and access to all clips refer to Schaefer et al.~\cite{Schaefer2016}. 

\section{Qualitative Results} \label{sec:results}

\subsection{Theme 1: Tactile Interaction as a Key Component of PARO's as an Emotion Regulation Tool} \label{sec:theme1}

Due to its animal-like physical attributes and movements, participants perceived PARO as a comforting companion. For instance, P01 likened PARO's presence to interacting with a real animal, finding comfort in its movements and companionship: \textit{"I could feel it move (.) so it was just like having a cat in my lap (.) which is something I find really calming and it looked like it was someone there with me watching the movie (.) which feels way less scary than watching it by myself}”. Engaging with PARO helped participants regulate their emotions by allowing attentional redeployment from parts of the movie they perceived as more stressful to the robot. P02 explained that when the robot engaged with them during the movie via its movements, it provided a positive distraction: "\textit{the happy wagging of the tail helped me to get to control my emotion [$\dots$] I was distracted basically [$\dots$] when I have the robot on my lap so then [$\dots$] I didn’t think much of the intense situation}”. Similarly, P03 highlighted how PARO's interactions diverted attention from stressors, offering a source of comfort: “\textit{I suppose because it was kind of like hugging in and it was very soft to feel that kind of almost cuddled up (.) It kind of broke my attention on the scenes and gave me something else to kind of focus on (.) I think that's what took out a lot of tension}”. 

\begin{figure}[ht]
  \centering
  \includegraphics[width=\linewidth]{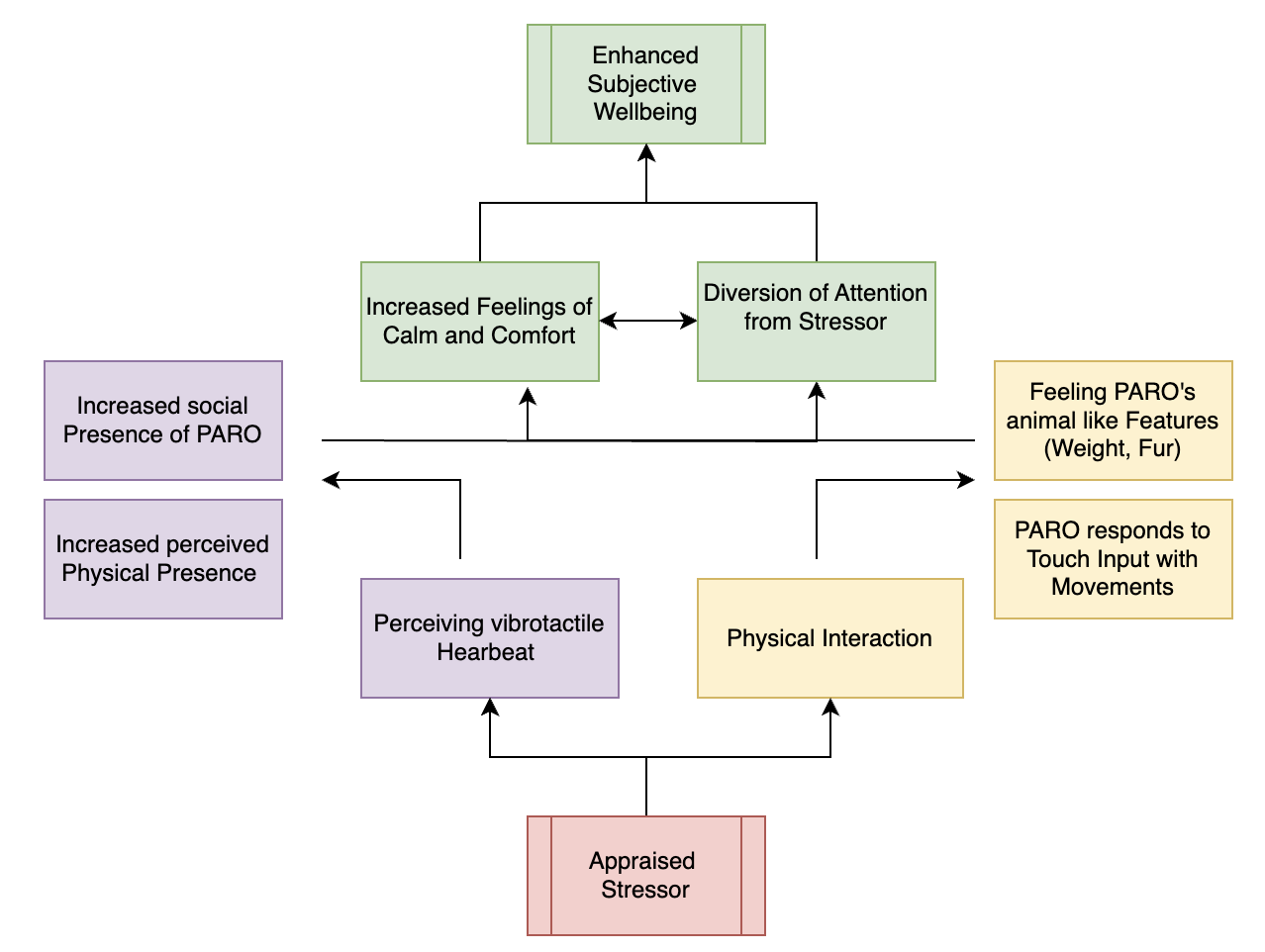}
  \caption{Visual representation of the impact of physical interaction with PARO and perceiving vibrotactile heartbeats on subjective wellbeing as discussed in Themes 1 and 2.}
    \label{fig:map}
\end{figure}

PARO's diverse interaction capabilities, such as head and flipper movements, provided comfort through tactile engagement: \textit{“it can interact with you (.) It can raise [$\dots$] his head and it move his tail and also [$\dots$] the flippers they would move from back and forth so it was very comforting}” (P18)\textbf{.} However, participants highlighted that it was not just about movement, but about PARO responding to their actions, leading to perceived reciprocal interaction. P04 explained how PARO's responsiveness to their touch fostered the illusion of responsiveness to their emotional needs:\textit{ "it’s like a positive feedback system where if I felt stressed out I would just like try to touch PARO more and it felt like it would react more to me touching it [$\dots$] it was actually responding to what I was feeling and that was a very nice feeling”}. However, perceived support from PARO was influenced by situational appraisal, with participants benefiting more during stressful scenes: \textit{“I feel like I didn't like pay much attention to PARO but later on like when it was more intense like I was definitely focusing on PARO more”} (P07), which highlights that engagement with the robot, and the influence it had on affect, was contingent on participants’ appraisal of the situation as threatening and their perceived need for support (Figure~\ref{fig:map}). 

Overall, PARO's animal-like physical attributes and interactive capabilities encourage tactile interaction, diverting attention away from the stressor and providing comfort to participants during the stressful moments of the video clips (see Figure~\ref{fig:map}). 

\subsection{Theme 2: Vibrotactile Heartbeats can enhance the Comforting Effects of PARO} \label{sec:theme2}

Several participants preferred PARO when augmented with a heartbeat, reporting feeling calmer due to the vibrations. For instance, P17 reported that PARO on its own primarily distracted them from the stressor, while PARO with the heartbeat provided comfort: \textit{“I feel like the first one [PARO with vibrations] comforted me but the second one [PARO] just made me remember where I was (.) because it kind of like brought me out of the situation”}. Similarly, P12 expressed they felt most calm when watching the clip with the vibrotactile heartbeats and noticed their absence when they interacted with PARO on its own: \textit{“Definitely think I felt calm (.) that was the most calm I felt of all of the clips and (.) and then Compared to the last one [PARO] I kind of miss like (.) I preferred having vibrations (.) just nice and comforting again”}. Similarly, P011 outlined that the robot with the vibrotactile stimulation provided comfort and emotional support in a similar manner to a pet: \textit{“I'm kind of afraid of these scary things so I need (.) A friend or pet or like a robot with the vibrations to help me to keep calm It can [$\dots$] And [$\dots$] I (.) I will not feel so scared when watch the movie\textit{”.}} For P08, the effects of PARO extended from the affect to their physiological stress response: \textit{“It felt like a heartbeat to follow almost (.) which sounds silly because I can't control my heartbeat [$\dots$] but I felt like it was uhm (.) Like my breathing was a bit more level because I was like following the kind of steady rhythm”}, showing that some users may synchronize their breathing to the rhythm of a vibrotactile stimuli, thus regulating their stress levels.

\subsubsection{Vibrotactile Stimulation Increases User Attention Towards the Robot} \label{sec:subtheme21}

Augmenting the robot with a vibrotactile heartbeat, a constantly pulsing stimuli, increased participant attention towards the robot. P04 explained that without the vibrotactile feedback they \textit{"could easily forget that it [PARO] was there"} whereas with the vibration they \textit{"were more actively aware that it was there".} Among others, P06 highlighted the importance of continually perceiving tactile stimulation when touching the robot to sustain the engagement, providing a constant physical reminder of reality and PARO’s companionship, which helped them to feel less affected by the stressor: \textit{"when you watch a movie you can have a tendency to kind of tunnel vision on what's going on and that can lead to like more exaggerated like an emotional response [$\dots$] with the vibrations like constantly It didn't matter if you weren't necessarily like moving your hand over the robot because the robot was reminding you that it was still there even without moving and which was a nice way to kind of prevent that like tunnel vision”.} Likewise, P01 highlighted that PARO without the vibrotactile feedback required more active input: \textit{"I was petting it to see if it would respond in any way"}, reducing perceived engagement, whereas PARO augmented with vibrations engaged their attention even when it was not producing movements, leading to increased awareness of PARO’s presence: \textit{“ it felt more grounding by itself even if I didn't interact that much with it [$\dots$]reminding me that the robot is there even when I'm not looking at it and I'm just looking on the screen [$\dots$] so I was more aware of it for sure”}. In summary, adding a vibrotactile heartbeat with a constant rhythm to the robot can reinforce its physical presence and redirect user attention. Users in our sample not only perceived this to increase their engagement with the robot, but also increase the positive impact of PARO as an emotion regulation tool. 

\subsubsection{Vibrotactile Heartbeat Increases PARO’s Social Presence and Life-likeness} \label{sec:subtheme22}

Augmenting PARO with a heartbeat enhanced the illusion of social presence. Some, such as P08, noted that the vibrations reminded them of real-life interpersonal touch, such as \textit{"lying on someone's chest or when you're hugging someone and you feel their heartbeat and you're like ‘oh This must be like what a baby feels when you're being like when they're being held"}. Similarly, P04 highlighted how vibrotactile heartbeats can evoke such experiences:\textit{ “feeling like some kind of heartbeat that reminds me of just being in in contact with another living being”}.

The added heartbeat led to more life-like touch experiences and greater social presence, with P06 describing the robot as \textit{“more alive in a way [$\dots$] you could kind of feel that you were in this together and watching this film which was quite a comforting thought”}. Some participants perceived the robot with a heartbeat to embody more of a social entity, which in turn increased the perceived socio-emotional support provided by the robot: \textit{“I feel that one understood me better [$\dots$] I felt more connected to it [$\dots$] I could feel it responding to my moods and also he [PARO] was more focused on me”} (P09). This was echoed by P10, who perceived the robot as more responsive and ascribed intentions and emotions to it as if it were a social entity:\textit{ “it feels like it's reciprocal (.) so previously [$\dots$] when I stroked it it did have some kind of response, but this time it felt as if it was also kind of properly enjoying it and trying to return that kind of that love or affection”}. Similarly, P05 reported that PARO with vibrations seemed more life-like and emotionally intelligent: \textit{“like it seemed like it was like ‘oh she's anxious’ like comforting and like uh moving a lot more (.) and be more responsive and everything (.) it's like more like a real animal”}, adding that the vibrations created the illusion of the robot intentionally touching the user: \textit{“it’s just like someone just touching [...] my hands”}.

Overall, participants expressed a preference for PARO augmented with a vibrotactile heartbeat, reporting feeling calmer and more engaged with the robot. The vibrotactile stimulation reinforced PARO's physical presence and redirecting user attention to it away from stressors, while also enhancing the perception of PARO as a social entity, increasing feelings of connection and emotional support (see Figure~\ref{fig:map}).

\subsection{Theme 3: Creating Effective Life-Like Vibrotactile Displays} \label{sec:theme3}

Individual differences across participants provided novel insights into the effective design of life-like vibrotactile displays. The appraisal of the haptic stimuli as a heartbeat determined the effect it would have on participants (see Figure~\ref{fig:map}). Participants were not explicitly informed that the vibrotactile condition represented heartbeats, reflecting an intentional choice by the researchers to assess whether this augmentation enhanced the robot's perceived life-likeness. Participants who identified the vibration as a heartbeat experienced an enhanced perception of the robot as discussed in Theme 2. However, a small group of participants did not identify the stimulus as a heartbeat, rather appraising it as a random vibration or even representative of an underlying mechanical process: \textit{“It was more like I'm touching something to collect my pulse or it's a measure or something"} (P03). This lack of meaning negatively affected the perception and interaction with the robot: \textit{"the vibrations at first felt a bit odd I would say and I think that kind of took away from some of my like initial feelings of it's kind of life-likeness”} (P10). These findings illustrate that merely incorporating vibrations into a robot does not automatically improve interaction. Instead, it is crucial for users to attribute meaning to the vibrations to fully benefit from them. This supports integrating vibrotactile displays into SARs, which evoke haptic cues people have experienced in the real world, such as haptic bio-physiological cues during close body contact with humans or animals. 

Moreover, the prototype hardware used, involving visible cables, was recognised by some participants as linked to the vibrotactile display. While most participants benefited from the display regardless of the hardware visibility, some found it challenging to associate the vibrations with the robot's behaviour due to the external attachment: \textit{"I feel like it was just an added thing that I was touching"} (P16). Similar to participants who perceived the vibrotactile cues as artificial due a lack of meaning, participants who perceived the cues as artificial due to the knowledge of them being artificially generated via the attachment felt it did not increase the robot's life-likeness: \textit{"It doesn't have any form of a living being (.) Yes it was just a little square that I had to hold and so I didn't see it as part of the robot”} (P13). Nonetheless, most users, despite knowing the artificial nature of the heartbeat, benefited from the vibrations as discussed in Theme 2. This suggests that, while participants reported a subjective shift in improved wellbeing and enhanced perception, refining vibration placement and enhancing spatialization could further improve these effects.

\section{Discussion} \label{sec:discussion}

\subsubsection{User Experiences with PARO during stressful settings}
During this study, we exposed participants to horror movie clips to understand if PARO was able to provide comfort during moments of stress. We found that they perceived PARO as a social substitute that offered companionship, comparable to a pet, which provided socio-emotional support in this stressful setting. While existing research extensively supports the positive influence of PARO on overall user wellbeing\cite{Wada2005,Aminuddin2017, Scoglio2019}, only a limited set of literature has delved into its specific effects during exposure to stressors~\cite{Borgstedt2022,Geva2020}. We contribute further evidence that PARO can assist users in managing emotions amidst mild to moderate stressors. Through conducting a qualitative study, we were further able to elucidate some of the underlying mechanisms that drove the positive effect in our sample. 

Participants either unintentionally shifted their focus onto the robot when they felt its movements or intentionally shifted their focus from the movie to PARO during high-emotion scenes. For those participants, PARO served as an active emotion regulation tool, allowing them to consciously divert their attention away from the stressor. Physical engagement with PARO, such as holding or touching it, played a pivotal role and offered a sensory distraction that contributed to emotional relief. However, the effectiveness of PARO's emotional support was contingent on participant appraisal of the situation as stressful or threatening, with most participants primarily interacting with PARO during stressful moments. These findings align with our previous findings, which demonstrated that users benefited from interacting with PARO only if they wished to be comforted but not if their goal was to complete a stressful task ~\cite{Borgstedt2022}.

Regarding the future use of PARO, 26/30~participants expressed a desire to incorporate PARO into their future experiences. Participants primarily envisioned using PARO during stressful or anxious situations, where they would benefit from socio-emotional support as a substitute for human social companionship. This preference aligns with existing research findings~\cite{Ihamäki2021,Chang2013,Hudson2020, Wada2007}, that underscore the potential therapeutic benefits of SARs. Only one participant explicitly stated their reluctance to consider future use of PARO, due to not finding it beneficial. Participants suggested using PARO primarily in home settings, but some also wanted to use it during stressful work-related tasks. However, usage of PARO in work-related contexts is dependent on individual user preferences, as six participants raised concerns about the robot being too distracting or inappropriate. It is important to note that the demographic composition of our study sample differs from that of most research in this field, which primarily focuses on assessing the usability of SARs among elderly populations~\cite{Wada2007,Hudson2020,Chang2013}. Almost all our participants were young adults, either engaged in academic pursuits or employed within a university context. This extends the potential applications of PARO beyond elderly and clinical populations, to a demographic less explored in research with touch-based SARs like PARO. 

\subsubsection{User Experiences with a Vibrotactile Heartbeat}
This study explored how augmenting a SAR with a vibrotactile heartbeat influenced people's responses to stressful situations. Overall, the majority of our participants preferred interacting with the robot when it was augmented with a heartbeat, with participants reporting increased feelings of calmness relative to using the robot without the vibrotactile heartbeat. While participants in this study and our previous work~\cite{Borgstedt2022} reported PARO to be a positive and helpful distraction, our findings show that PARO on its own was not sufficient to comfort users when their attention was directed towards a stressor, as PARO’s interactive behaviours are dependent on user input, which was reduced when users were focused on the ongoing stressor. Instead, our findings provide novel evidence by showing that augmentation with a vibrotactile heartbeat, diverted the user's attention more effectively from the stressor onto the robot, which helped users to feel more comforted and in turn increased the perceived socio-emotional support provided by PARO. These findings align with psychological evidence from~\cite{Janson2017}, who suggest that effective cognitive distraction strategies are associated with decreased subjective and physiological stress responses. Our findings show that augmenting a SAR with vibrotactile heartbeats, and thus diversifying its pyhsical interaction modalities, increases the effectiveness of the SAR as a cognitive distraction strategy during stressful situations. These findings have important implications for the HRI community, as they show that creating companion robots that aid users in regulating their subjective mood and wellbeing during stressful situations, necessitates the development of physically engaging interaction modalities, which effectively divert the users attention from the stressor onto the external environment. The study's findings underscore the significance of augmenting SARs with vibrotactile heartbeats in enhancing users' comfort and reducing stress levels, highlighting a promising avenue for HRI research to foster increased socio-emotional support through the expansion of touch-based interaction modalities. 

The vibrotactile heartbeat also had a positive influence on user perceptions of the robot, as they frequently reported that the robot was more active, lively, and responsive to their behaviour and even emotions. Overall, the vibrations increased PARO’s social presence by simulating life-like haptic experiences. Participants felt as though they were in contact with a living being, leading to increased emotional responsiveness and perceived socio-emotional support. These findings are related to those from~\cite{Nie2012}, who found that augmenting a robot hand with more life-like warmth was associated with increased feelings of friendship and trust toward the robot. Similarly, Yohanan et al.~\cite{Yohanan2011role} found that simulating life-like physiological signals such as ear twitching or breathing, encouraged users to engage in affective social touch and emotional communication with a SAR. These studies highlight that incorporating life-like and varied haptic experiences into physical human-robot interaction can enhance the perception of and relationship with a SAR.

Finally, the user experiences gathered in this study provided insights to inform the optimisation of life-like, vibrotactile stimuli for SARs, and specifically a vibrotactile heartbeat. In this study, to avoid priming participants, they were not told that the vibrations represented a heartbeat. Interpretation of the vibration’s purpose and nature influenced its impact. Those who did not recognise the nature of the vibration, perceiving it as random or mechanical, did not necessarily find it comforting as they often viewed the vibrations as a by-product of a technological process. This finding underscores the importance of creating stimuli that evoke the user’s prior emotional experiences, as discussed in~\cite{MacDonald2020, Marchetti2022}. The presentation of the vibrotactile actuator influenced some participants’ perceptions. The vibrotactile patch was attached to the robot and, despite being camouflaged with fur, some cables were visible, reducing the illusion that the heartbeat was coming from within the robot. While most participants overlooked it and focused on the vibrations, a small group of users struggled to associate the vibrations with the robot. This provided important insights into the development of vibrotactile features, emphasising for them to emanate from within the robot.

\section{Conclusion}
In this study, we investigated the impact of augmenting a PARO robot with a vibrotactile heartbeat on users' subjective wellbeing during exposure to an emotional stressor (a horror movie). Overall, the PARO robot provided socio-emotional support to users during the stressful moments of the horror movies, through its interactive capabilities and animal-like physical features. We found that augmenting a PARO with a vibrotactile heartbeat, which emulates haptic physiological responses of a real animal, enhances PARO's effectiveness as as an emotion regulation tool and social companion. Most users in our study successfully identified the vibrotactile stimulus as a heartbeat, and reported the robot to be more life-like, socially present, and providing more socio-emotional support compared to without the heartbeat. Our findings underscore the importance of iterative prototyping life-like haptic cues together with prospective users, as the positive impact of the vibrotactile heartbeat stimuli was influenced by the ability to correctly identify the stimulus as a heartbeat and the perceived realism of the heartbeat. By gaining a deeper insight into users' positive experiences with the heartbeat and its limitations, we provide valuable guidance for the design and implementation of life-like vibrotactile cues for zoomorphic SARs like PARO.

%% BIBLIOGRAPHY
\bibliographystyle{IEEEtran}
\bibliography{bibliography}

\end{document}